# A RISC-V Simulator and Benchmark Suite for Designing and Evaluating Vector Architectures

CRISTÓBAL RAMÍREZ, Polytechnic University of Catalonia and Barcelona Supercomputing Center
CÉSAR A. HERNÁNDEZ, National Polytechnic Institute of México and Barcelona
Supercomputing Center
OSCAR PALOMAR and OSMAN UNSAL, Barcelona Supercomputing Center
MARCO A. RAMÍREZ, National Polytechnic Institute of México
ADRIÁN CRISTAL, Polytechnic University of Catalonia and Barcelona Supercomputing Center

Vector architectures lack tools for research. Consider the gem5 simulator, which is possibly the leading platform for computer-system architecture research. Unfortunately, gem5 does not have an available distribution that includes a flexible and customizable vector architecture model. In consequence, researchers have to develop their own simulation platform to test their ideas, which consume much research time. However, once the base simulator platform is developed, another question is the following: Which applications should be tested to perform the experiments? The lack of Vectorized Benchmark Suites is another limitation. To face these problems, this work presents a set of tools for designing and evaluating vector architectures. First, the gem5 simulator was extended to support the execution of RISC-V Vector instructions by adding a parameterizable Vector Architecture model for designers to evaluate different approaches according to the target they pursue. Second, a novel Vectorized Benchmark Suite is presented: a collection composed of seven data-parallel applications from different domains that can be classified according to the modules that are stressed in the vector architecture. Finally, a study of the Vectorized Benchmark Suite executing on the gem5-based Vector Architecture model is highlighted. This suite is the first in its category that covers the different possible usage scenarios that may occur within different vector architecture designs such as embedded systems, mainly focused on short vectors, or High-Performance-Computing (HPC), usually designed for large vectors.

New article, not an extension of a conference paper.
This work is partially supported by CONACyT Mexico under Grant No. 472106 and the DRAC project, which is co-financed by the European Union Regional Development Fund within the framework of the ERDF Operational Program of Catalonia 2014-2020 with a grant of 50% of total cost eligible.
Authors' addresses: C. Ramírez, C. A. Hernández, O. Palomar, O. Unsal, and A. Cristal, Barcelona Supercomputing Center, Carrer de Jordi Girona, 29,31, Barcelona, Spain, 08034; emails: {cristobal.ramirez, cesar.hernandez, oscar.palomar, osman.unsal, adrian.cristal}@bsc.es; M. A. Ramírez, Computer Research Center, National Polytechnic Institute of México, Av. Juan de Dios Bátiz s/n, Nueva industrial Vallejo, Gustavo A. Madero, Mexico City, Mexico, 07738; email: mars@cic.ipn.mx.

## 1 INTRODUCTION

During the past four decades, the evolution of supercomputing has been very progressive. In 1997, the fastest computer in the world, the ASCI Red featuring Intel processors, had a top performance of 1.06 TeraFLOPS, while in 2020, a single Intel processor tops 1 TeraFLOPS. In fact, Exascale computing has become the new milestone for supercomputing. In the most basic sense, Exascale (1,018 floating-point operations per second) will provide the capability to perform more realistic simulation about the processes involved in precision medicine, regional climate, the unseen physics in materials discovery and design, the fundamental forces of the universe, and much more. However, to arrive at Exascale levels, architectural innovations, technology breakthroughs, and hardware and software coordination are needed. In this sense, different nations have increased their investments in HPC projects such as the European Processor Initiative (EPI) [1] in Europe, the Exascale Computing Project (ECP) in the US [2], the Post-K project in Japan [3], and the National Key R&D project in China [4], competing in what today is called "The race to Exascale."

Parallelism at multiple levels is now the driving force of computer designs where energy is one of the primary constraints. One effective way to achieve high performance and efficiency is the exploitation of data-level parallelism (DLP). In this sense, parallel architectures can deliver good performance at a lower cost. One category of parallel hardware organization is termed Single Instruction Multiple Data (SIMD) [5]. Two variants of SIMD are multimedia extensions and vector architectures [6]. Multimedia extensions allow executing a set of predefined operations over vector registers of a fixed length. In contrast, in a Vector Architecture, there is no single preferred vector length, just the Maximum Vector Length (MVL) is defined, and the application can use any vector length that does not exceed the maximum. Nowadays, most commodity CPUs implement architectures that feature SIMD instructions. Common examples for Multimedia extensions include Intel x86's MMX, SSE, AVX, AVX2 and AVX-512 [7], MIPS's MSA [8], ARM's NEON [9]; for vector architectures, the well-known vector extensions for NEC [10] and CRAY, and the new vector extensions ARM's SVE [11] and RISC-V V extension [12], which are considered part of "*the reemergence of vectors.*"

Exascale systems will be strongly constrained by energy efficiency. In that sense, SIMD processing plays an important role in the development of the new Exascale systems. SIMD instruction set architectures (ISAs) are very expressive, enabling a good representation of high-level HPC algorithm semantics to be carried into the microarchitecture. Also, SIMD is potentially more energy-efficient, since a single instruction can launch many data operations. Finally, and perhaps the biggest advantage of SIMD from the software perspective, is the ease-of-programmability, since the programmer continues thinking sequentially, and SIMD is close to the already familiar Single Instruction Single Data (SISD) category.

In that sense, RISC-V [13] ISA is opening new opportunities for academia and industry with the incorporation of the new vector extension. In fact, this vector extension has arrived just at the most convenient moment where the quest for extreme energy efficiency hardware has renewed interest in vector architectures. It has not been long, since RISC-V announced the first stable release of the RISC-V vector extension, and there are already several open and commercial-based products. Some examples are Ara [14] from ETH Zurich and Xuantie-910 [15] from the Chinese company

Alibaba, to name a couple. Also, in the HPC projects mentioned above, clear research examples on vector architectures can be seen. In the case of the EPI project, a design based on the new RISC-V vector extension was proposed as one of the essential points to develop power-efficient and high-throughput accelerators. In the case of the Post-K project, breaking news puts Japan at the top in the TOP500 [16] list released in June 2020. Fujitsu put into operation parts of the Fugaku supercomputer originally scheduled to start operating in 2021. Fugaku is built with the Fujitsu A64FX microprocessor based on the ARM 8.2-A architecture. This architecture adopts the Scalable Vector Extension (SVE) as an efficient way to achieve Exascale computing. Since there is no single solution to arrive at Exascale, novel ideas on vector architectures must be explored.

### 1.1 Tools for research on Vector Architectures

One of the most-used platforms for computer-system architecture research encompassing system-level architecture as well as processor microarchitecture is gem5 [17], which can be used to test those novel ideas on vector architectures. For Multimedia extensions, gem5 has support for the Intel's MMX and SSE (64-bit and 128-bit extensions), which are implemented as part of the core microarchitecture. However, the support for the more current extensions such as AVX2 and AVX-512 is missing. On the vector architecture side, there is full support for the ARM SVE, where the MVL allowed by the architecture is 2,048-bit (32 elements each 64-bit). However, despite the relevance of vector architectures, gem5 does not have a public distribution, which includes a vector architecture model that evaluates different implementations including short (around 512-bit), medium (around 4,096-bit), and large (16,384-bit or more) vectors combined with a flexible and customizable model that fits with the research requirements. In consequence, researchers have to limit their explorations to the MVL allowed by the current models, which decreases the possible scenarios that could allow a flexible and customizable model without MVL limitation; either, researchers have to develop their own environment to test their ideas, which is very time-consuming. In this sense, the incorporation of the new RISC-V vector extension will offer to the computer architecture community maximum freedom in the research and development of new acceleration technologies where the MVL can be chosen by the architect instead of being restricted by the architecture.

On the benchmark suites side, as novel architecture designs have appeared, the need for new benchmark suites arises. There are several suites to measure single-core performance over data-parallel applications such as Parboil [18] and Polybench [19]. Also, there are several suites focused on parallel computing on general-purpose CPU architectures such as PARSEC [20] and HPC Challenge Benchmark Suite [21], as well as others for heterogeneous computing such as Rodinia [22] and Polybench/GPU [19], covering MPI, OpenMP, OpenCL, and CUDA programming models, while SIMD Suites are very limited, such as ParVec [23]. It is well-known that many applications can benefit from vector execution achieving higher performance, higher energy efficiency, and greater resource utilization. However, the effectiveness of the hardware depends not only on the design but also on the compiler's ability to vectorize the code to be executed. As reported in Reference [6], there is a tremendous variation in how different compilers perform in vectorizing programs. Supporting auto-vectorizing large codes is currently too limiting. Relying on good performance typically relies on the programmer's effort; for example, rewriting to obtain well-structured control flow or vectorizing the code using intrinsics. This effort is one of the principal reasons for not having many vectorized benchmark suites. Despite this, suites to evaluate the different modules that compose a Vector Architecture have received little attention from previous work on benchmark development.

The contribution of this article is to present a simulator and benchmark framework, which enables researchers to test novel ideas on vector architectures. Our gem5-based simulator baseline model corresponds to a decoupled vector architecture, and different vector micro-architecture implementations can be evaluated, since the number of physical vector registers, MVL, number of queue entries, issue scheme, number of lanes, the latency of the functional units, latency and topology of the lanes interconnection, and number of memory ports are customizable. To help evaluate these architectures, a novel Vectorized Benchmark Suite was developed that covers the different possible scenarios that may occur within different vector architecture designs that can operate from short MVL to large MVL, taking into account the different modules that can be evaluated in a vector architecture such as the lanes, the interconnection between lanes, and the memory management.

This article is organized as follows: In Section 2, the background and some academic and industrial efforts on vector architectures is given. In Section 3, a detailed description of our vector architecture model implemented on gem5 is shown. Then, the RISC-V Vectorized Benchmark Suite is presented in Section 4, describing how the vectorized versions were implemented and showing the degree of vectorization achieved. Once both tools are detailed in Section 5, a study of the scalability for each application executed on different configurations of the gem5-based vector architecture model is highlighted. Section 6 focuses on the related work. Finally, Section 7 summarizes the key points of this work and gives some examples of next-generation ideas.

## 2 BACKGROUND

This section presents some important concepts of vector architectures, which are needed to better understand the Vector Architecture model presented in Section 3. Also, some important academic and industrial efforts on vector architectures are shown.

### 2.1 Vector Architectures

An elegant interpretation of SIMD is called a Vector Architecture, which has been closely identified with supercomputers designed by Seymour Cray. A key element of these architectures is that arithmetic/logic and load/store instructions operate on sets of vectors instead of individual data items. Moreover, instead of having, for example, 32 Arithmetic Logic Units (ALU) to perform 32 operations simultaneously, vector architectures typically pipeline the ALU to obtain a good performance at a lower cost. One of the main features of vector architectures is the Vector Register File (VRF), where each vector register can hold a large number of elements, and the maximum number of elements are represented by the MVL parameter, which can vary depending on the hardware implementation [6].

Vector architectures that include multiple parallel pipelines, also known as lanes, can produce two or more results per clock cycle. Adding multiple vector processing lanes is a popular technique that leads to an advantage in performance and scalability, as shown by Asanovíc [24]. In a multi-lane vector architecture, one lane operates with a register subset of the overall VRF and a data path subset of the overall vector functional units data paths, where all the lanes work fully synchronized [25]. Furthermore, multi-lane vector architectures need extra hardware to control the synchronization between lanes and also a lane-interconnection network to allow data movement between all the lanes.

Vector architectures have been traditionally applied to the supercomputing domain (Cray, NEC, IBM) in the 1970s up to the 1990s. The Cray-1 [26] introduced in 1976, was a register-based machine and the first supercomputer to successfully implement the vector processor design where arithmetic instructions operate on vector registers while separate vector load and store instructions move data between memory and vector registers. In the early '80s, Japanese manufacturers (NEC, Fujitsu, and Hitachi) entered the vector supercomputer market, introducing lines of parallel vector computers [27]. In the early '90s, there was a radical change in the computer industry. The introduction of faster microprocessors substantially changed the supercomputing market mainly because the FLOPS/$ is substantially lower for commodity-based supercomputers, although vector supercomputers could achieve higher FLOPS. Thus, the idea of building parallel machines based on many out-of-order microprocessors offered an attractive alternative instead of vector supercomputers [27].

In the late 1990s, there were many academic research proposals on vector architectures. Espasa [28] proposed using decoupling techniques in a vector processor by splitting the instruction stream into three different streams through a set of queues: scalar computation instructions, vector computation instructions, and memory accessing instructions (both vector and scalar), and showed that the performance of vector programs could be significantly improved. In a second study, Espasa [29] demonstrated that dynamic scheduling commonly applied to the superscalar processors such as register renaming and out-of-order execution could also be applied to the vector processors and obtain significant advantages. Applying dynamic scheduling, they reported a speedup of 1.24–1.72x for realistic memory latencies. Those academic ideas were studied by the processor manufacturers to conceive new designs. A clear example is the Cray X1 [30] launched in 2003, which is a distributed shared memory multiprocessor with a vector ISA (NV-1), capable of scaling to thousands of processors. The design features a VRF that holds 32 physical registers with an MVL of 64 elements where each element is 64-bit. The Cray X1 was the first to implement a decoupled vector micro-architecture. Decoupling between the vector memory unit and the vector execution unit facilitates the dynamic tolerance of memory latency. Moreover, decoupling between the vector and scalar execution units allows scalar execution to run ahead. The next Cray design launched in 2007 was called BlackWidow [31]. Like its predecessor, BlackWidow implements decoupling from the scalar core and improves scalar-vector synchronization primitives with a new vector ISA (NV-2). A large VRF was implemented that has 32 physical registers with an MVL of 128 elements each 64-bit wide. The design was organized as an eight-lane configuration, where each lane is associated with 16 elements of every vector register.

One modern example of a vector architecture is the SX-Aurora TSUBASA [10, 32], launched in 2018. SX-Aurora TSUBASA features eight vector cores in a single chip with a frequency of 1.6 GHz. Each vector core includes a scalar processor that provides the basic functionality as a processor (Fetch, decode, exception handling, etc.) and a decoupled Vector Processing Unit (VPU). The VPU includes renaming with 256 physical registers and Out-of-Order scheduling. The MVL is 256 elements each 64-bit wide, and it has 32 Vector Lanes with each lane featuring four pipelines (FMA0, FMA1, ALU0/FMA2, and ALU1/Store), executing up to three arithmetic operations plus one memory operation in parallel.

## 3 GEM5—VECTOR ARCHITECTURE MODEL

The gem5 simulator has been extended to model a decoupled vector architecture. The customization provided by the parameter-based model allows the designer to obtain a vector engine design capable of achieving a tradeoff between performance, energy efficiency, and area. In that sense, it is possible to simulate a design that fits with the researcher requirements. For example, a vector engine designed for HPC, by setting a design for large vectors (256 64-bit elements), composed of a renaming unit capable of supporting 64 physical registers, a vector arithmetic and a memory queue with 16 entries; these features can be set up to work with, say, eight lanes. In contrast, the vector engine can also be targeted for the embedded market segment by setting a design for short vectors (8 64-bit elements), reducing the number of available physical registers, and with only one-lane configuration. The main goal of this work is to obtain the more flexible and customizable vector engine for researchers; in that sense, several decisions were taken into account.

**Create a model based on RISC-V.** The adoption of RISC-V is a key factor for having the maximum freedom both in research and development of the new technologies, without the limitation from hardware and software ecosystems. Furthermore, the new V extension includes a key feature that is unbounded MVL size, which, in combination with a flexible and customizable model, could lead to set designs that support very long vectors.

**Decoupled vector engine.** A decoupled design provides several advantages. First, decoupling between the vector memory unit and the vector execution unit facilitates the dynamic tolerance of memory latency. Second, decoupling between the vector and scalar execution units allows scalar execution to run ahead. Implementing a vector engine as a pipeline tightly coupled to an aggressive out-of-order superscalar core is a typical implementation. In fact, most commodity CPUs that feature SIMD instructions work in this way. However, these designs are optimized for short vectors such as Intel's AVX-512, where the VRF size is around 2 KB (32 vector registers each 512-bit wide), and the area overhead added to the superscalar core is acceptable, without limiting the maximum frequency of the superscalar core. The case for vector engines designed for long vectors is a different story. In this case, the VRF size is around 64 KB (32 vector registers each 16,384-bit wide) or more when implementing renaming, which could lead to a big area overhead and could limit the maximum frequency of the superscalar core. Contemporary vector architectures are implemented as decoupled engines running around 1.5 GHz (e.g., Cray BlackWidow runs at 1.3 GHz, and SX-Aurora TSUBASA runs at 1.6 GHz), mainly limited by the big structures needed to hold long vectors. In contrast, superscalar cores run at higher frequencies. Furthermore, by having a decoupled design, it is possible to study different possibilities to get energy-efficient architectures. Examples include clock gating, turning off the clock of inactive modules to save energy and dynamic power; or applying dynamic voltage-frequency scaling when there are periods of low activity where there is no need to operate at the highest clock frequency and voltage.

### 3.1 Scalar Core

Different fully parameterizable CPU models are provided by gem5, such as the in-order CPU and the out-of-order CPU, which allows micro-architectural simulations. In this work, extra support to the in-order CPU pipeline was added to recognize the vector instructions and perform different tasks before sending these instructions to the vector engine. The core runs concurrently with the vector engine, so most of the scalar operations are amortized underneath vector execution. The scalar core is responsible for fetching and decoding the vector instructions and carrying it through the pipeline. Most of the vector instructions are treated as a *nop* operations in the scalar core. Furthermore, if the vector instruction has a scalar operand as a source, it must read the scoreboard to check if the source operand is ready. Then, the vector instructions continue to the next stages until they reach the commit stage, where they are sent to the vector engine. With this, the vector instruction execution will not be interrupted by any possible control hazard, such as a miss branch prediction generated by older scalar instructions.

One of the current limitations is that RISC-V is supported only in Syscall Emulation mode, which implies that interrupts, exceptions, and fault handlers are trapped and managed by the host without running a handler routine to manage the event. More detailed information about current limitations is given in Section 3.2.6. Having said the above, once the vector instruction is sent to the vector engine, it can be retired from the scalar pipeline, since any exception generated by a vector instruction, such as a page fault caused by a memory request, is trapped and managed by gem5.

### 3.2 Vector Engine

Once the vector instructions arrive at the vector engine, they are first renamed to remove the false dependencies, increasing the amount of instruction-level parallelism (ILP) that can be exploited. Then, two operations are performed in parallel. The first is to assign one entry in the reorder buffer, and the second is to allocate the instruction in a temporal queue depending on the instruction type (arithmetic or memory). Once assigned in the corresponding queue, the instruction waits until it fulfills the requirements to be issued; its operands become ready, and the corresponding execution unit is available (the memory unit or the vector lane). Then, when the instruction completes execution, the commit is made by retiring the instruction and freeing up the hardware resources consumed. Our decoupled architecture design has some other unique features for efficiency. For example, as explained in Section 3.2.3, the vector lane architecture was carefully tuned to minimize pipeline bubbles due to structural hazards. The following subsections present a more detailed description of the components of the vector architecture model.

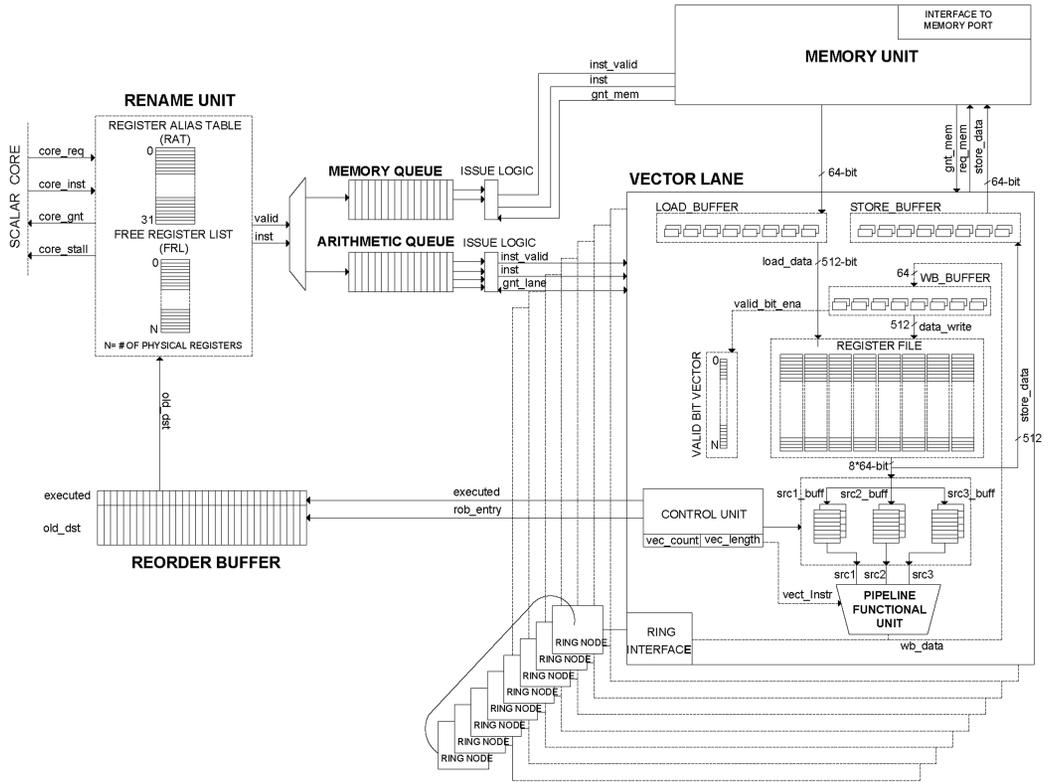

Fig. 1. gem5 Vector architecture model.

Figure 1 shows the general view of the vector engine model. Some specific configurations are also included to explain the interaction between the internal modules. The model presented corresponds to a multi-lane vector engine. By setting eight lanes, only one memory port could be enough to feed all the lanes, taking into account that the cache line size is set to 512-bit (8 elements each 64-bit), and with every cache line request it is possible to send one element to each lane in an interleaved fashion. The MVL is set to 16,384-bits (256 elements each 64-bit). The VRF line size is set to 512-bits. Also, a ring topology for lane interconnection is chosen.

*3.2.1 Vector Renaming.* As part of the dynamic scheduling implemented in the design register renaming is performed. The goal of the renaming is to remove false dependencies by changing the names of the source logical registers to its corresponding physical register that was mapped previously. Additionally, the logical destination register is renamed to a new physical register. This is performed by reading a structure termed as the Free Register List (FRL), which contains all the available physical registers. This mapping is stored in another structure termed as the Register Alias Table (RAT), where the logical destination operates as the write address. At the same time, the logical sources and destination registers read the RAT to obtain the corresponding physical source registers and the physical destination register that was mapped by a previous instruction, also known as *old-destination*. Then, these structures are coupled with a dependency check logic to analyze the instruction and solve any write-after-write dependencies. The physical registers in-flight (*old-destinations*) that are no longer used are appended to the free register list at commit time. Detailed information about the commit process can be found in Section 3.2.2. Finally, the number of physical registers can be set by the designer.

*3.2.2 Reorder Buffer.* The implemented vector architecture model permits choosing the issue scheme, which can be in-order or out-of-order; for that reason, a structure to preserve the program order is needed. A Reorder Buffer (ROB) is a structure that allows instructions to be committed in-order. Also, it holds important information about the instruction that can be useful during and at the end of its execution, such as the program counter, the physical *old-destination*, and a bit field termed as *executed*. The *executed* bit allows to know if the instruction has been completed or not. The number of ROB entries can be set by the designer.

When a new instruction arrives at the ROB, it is allocated in the next available entry signaled by the tail pointer (write pointer). At the same time, the address of the assigned entry is sent together with the instruction to the corresponding queue. In that way, the instructions know their locations in the ROB, and they can write to it when it is needed.

When an instruction finalizes its execution, the *executed* bit field associated with the corresponding ROB entry is set. It means that the instruction is ready to be committed. However, since the commit is performed in-order, the instruction must wait its turn to start this process. The head pointer defines the turn (read pointer). When the instruction pointed by the head pointer has the *executed* bit set, it means that it can commit. If this is the case, the physical *old-destination* is written back to the FRL structure, to be assigned later to a new instruction. Also, the head pointer advances to the next entry to evaluate a new instruction in the next cycles.

*3.2.3 Vector Issue Queues.* As mentioned before, the design of the vector engine corresponds to a decoupled vector architecture, meaning that memory instructions and arithmetic instructions are buffered in different queues (Arithmetic Queue and Memory Queue) until fulfilling all the requirements to be issued. In this scheme, it is allowed to execute independent memory instructions ahead of arithmetic instructions and vice versa. This stage is called Issue, and it is in charge of dispatching instructions to the vector lanes or to the vector memory unit.

The issue stage is composed of two fundamental modules that are termed as Instruction Queue and the Issue Logic. The scheduling can be configured to use an in-order or out-of-order issue scheme. In addition, the number of entries in the issue queues also can be configured.

The instructions are issued as soon as they fulfill the requirements. First, the source operands must be ready; this is done by reading a structure called *Valid-bit* (more detailed information about the *Valid-bit* structure can be found in Section 3.2.4). Second, the hardware resources needed for execution must be available. An important restriction is that the vector lanes only support the execution of one arithmetic instruction at a time. This means that for certain arithmetic instruction, all source operands can be ready. However, the issue queue must wait until the previous instruction finishes its execution. Note that it is possible to execute memory operations at the same time.

In the special case of the memory queue, if an out-of-order issue scheme is selected, a dynamic memory disambiguation logic is enabled to check for possible memory hazards between load and stores held in the queue. Once the instruction arrives at the memory queue, the disambiguation process sets a bit called *memory hazard*. First, the load is disambiguated against all the stores in the queues. In this case, for every memory reference (load/store), there is a Vector Base Address (VBA), a Vector Length (VL), a Vector Stride (VS), and Standard Element Width (SEW) in bytes. The memory range accessed by a vector reference is defined as a set of memory locations located between VBA and VBA+(VL*VS*SEW) -1. Then, there is a memory hazard between a vector load and vector store if their corresponding memory ranges overlap at least one byte. Scatters/gathers operations (more detailed information about gather/scatter instructions can be found in Section 3.2.5) represent a special case where characterizing by a memory range implies more complex implementations. Then, these operations are executed in order.

Fig. 2. VRF elements distribution for a MV=256 elements and eight-lane configuration.

*3.2.4 Vector Lanes.* Figure 1 shows a simplified picture of the internal modules that comprise one vector lane. The vector engine can be configured with the required number of lanes. A key aspect is the VRF. In gem5, this is modeled as a simple memory, and it is possible to choose the number of read/write ports. However, the designer should take into account that in a hardware implementation, the number of ports would be highly constrained mainly in a large register file. This is because adding additional ports to an SRAM memory could lead to an increase in area and also could limit the maximum operating frequency or require more than one cycle to read/write in the VRF.

One important source of overhead is the *start-up time*, which is the latency in clock cycles until the pipeline is full [6]. The start-up time is principally determined by the pipeline latency of the vector functional unit. Moreover, the number of read ports in the VRF also can influence the *start-up time*. For example, in a vector engine designed for low power, a one read/write port SRAM memory can be used. In that sense, when a vector multiply-add operation arrives at the lane, in a first cycle, it can read the source1, in a second cycle, it can read the source2, and finally, in the third cycle, it can read the source3. All these read operations take three cycles, which are added to the start-up time. On the contrary, if a VRF with three read ports and one write port is chosen, the read of the three operands can be made in only one cycle. It is a design decision that can be taken according to the final target. For large vectors where it takes several cycles to execute an arithmetic operation, paying three cycles could be negligible. For a short vector where the full vector can be computed in less than a dozen cycles, paying three cycles in every instruction could lead to a serious loss of performance.

When multiple lanes are enabled, each lane operates with a register subset of the overall VRF. The elements of a vector register are interleaved across all the lanes. Figure 2 shows a detailed example using the same configuration presented in Figure 1 (eight-lane configuration with an MVL of 256 elements). Lane 0 is the owner of element 0, lane 1 is the owner of element 1, lane 2 is the owner of element 2, and so on.

The designer can specify the VRF line size (512-bit in the example shown in Figure 1). Then, every read operation to the register file will return a VRF line size; for that reason, it is necessary to perform operands buffering to store the elements read and to keep a constant stream of data to the functional unit, avoiding bubbles in the pipeline.

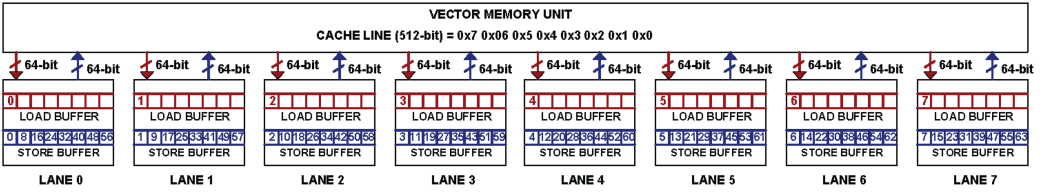

Fig. 3. Vector load/store buffer behavior.

As soon as the first result is computed, it is sent to a structure called *Write-back Buffer* (WB). This structure holds the resultant data (one 64-bit element per cycle) from the functional units. Once the WB buffer gets the total elements corresponding to one VRF line (512-bit in the example shown in Figure 1), the data can be written back in the VRF.

Each lane has a 64-bit bus to communicate with the Vector Memory Unit. When a load operation is performed, the Vector Memory Unit receives a complete cache line (512-bit in the example shown in Figure 1) and sends it in parallel to each lane (one 64-bit element). The Load Buffer (LB) is the structure in charge of collecting data from the Vector Memory Unit. Once the LB completes one VRF line (eight 64-bit elements), it proceeds to write back to the VRF. Figure 3 shows a more detailed example with the same configuration presented in Figure 1. In the vector memory unit side there is a cache line with eight 64-bit elements. Those elements are sent in interleaved fashion to each lane, meaning that Lane 0 is the owner of element 0, lane 1 is the owner of element 1, lane 2 is the owner of element 2, and so on. On the contrary, store operations read a complete line (512-bit) from the VRF and store it in the Store Buffer. This operation is performed at the same time in all lanes. Then, the store buffer sends one 64-bit element to the Vector Memory Unit each cycle.

When an instruction completes execution, the corresponding physical destination must be marked as ready to issue new instructions that were waiting for it. This is done in a structure called *Valid-Bit*, which for every physical register, one bit is added to the structure. For example, for a vector engine with 64 physical registers, a 64-bit Valid bit structure is implemented.

*3.2.5 Vector Memory Unit (VMU).* VMU receives instructions (load/store) from the memory queue, and it cannot accept a new instruction until it finishes its current work. This module is in charge of managing the requests to memory. The VMU supports unit-stride, strided, and indexed (gather/scatter) addressing modes. Vector unit-stride operations access elements stored contiguously in memory, starting from the effective base address. Vector strided operations access the first memory element at the effective base address and then access subsequent elements at address increments given by the byte offset specified by a scalar source. Vector-indexed operations add the contents of each element of the vector offset operand specified by the second vector source operand to the effective base address to give the effective address of each element.

Once it receives the memory instruction, as well as the memory address, the VL, and the stride (1 for unit-stride access), the VMU generates all the requested addresses and puts them in a FIFO. Then, the requests to memory are sent in-order in a pipeline fashion. However, the memory system can answer in a different order (hit under miss support). Additionally, it is possible to set the number of Miss Status and Handling Registers (MSHRs), which for a large vector implementation could be a relevant factor to consider. The MSHRs implement a queue that holds the list of outstanding memory requests. Each memory request is assigned to an MSHR object that represents a particular block of memory that has to be read or written to complete the command. When the memory request is sent, a unique order number is assigned to each read/write request as they appear on the slave port.

The memory port can be connected directly to any level of the memory hierarchy, unlike the Vector Lane where the timing is defined by the vector engine model. Once a request to the memory system is sent, the timing is managed by the gem5 memory model. There are several configurations that can be used; for example, it is possible to connect the vector memory port directly to the L1-Data cache. In another configuration, it may be possible to bypass the first level cache. It is normal that these kinds of architectures with large vector lengths are connected directly to L2 cache, since a vector memory instruction can amortize long memory latency over many elements with a small performance degradation [33].

*3.2.6 Lane Interconnection.* The vector extension can be configured with different numbers of lanes, where the lanes work fully synchronized. However, there is a class of instructions that involves communication between different vector lanes, basically for moving and addressing data such as vector *slides, vector reductions,* and vector *register gather* instructions. The *slide* instructions move elements up and down a vector register. The vector *reduction* instruction takes a vector register group of elements and performs a reduction using some binary operator to produce a scalar result that is written in the element 0 of a vector register. The *vector register gather* instruction reads elements from a first source vector register group at locations given by a second source vector register group and writes it in a destination vector register.

Therefore, an interconnection network is necessary to support this class of instructions. Two different interconnection networks (crossbar and ring network) are modeled. In the example shown in Figure 1, the vector lanes include a ring node (router) to communicate with the neighboring node. This interconnection could limit the performance for those applications that make intensive use of the lane interconnection, but it is cheap in terms of area. On the contrary, the crossbar interconnection could achieve an excellent performance, but it implies a considerable increase in area.

*3.2.7 Capabilities and Limitations.* Two different simulation modes are provided by gem5, Full System (FS) mode, and Syscall Emulation (SE) mode. The first provides the ability to simulate a full system. It can boot an operating system, handle interrupts, exceptions, and fault handlers. The second, the SE mode, focuses on the CPU and memory system and does not emulate the entire system. Syscalls are emulated, typically by calling the host OS. The gem5 RISC-V implementation still does not have the support to run in FS mode. Consequently, the vector architecture model is available only to run in SE mode.

Regarding the RISC-V vector extension, three versions (0.7, 0.8, 0.9) have been released so far. However, between the three different versions, the changes are small. For sure, there will be more updates before the specification is frozen as an official release, and it is believed that point is close. In that sense, this work has started to add vector extension support to gem5. However, the full specification is not implemented, leaving as future work the implementation of atomic operations, permutation operations, register grouping, and exception handling for the full system mode.

*3.2.8 Early Access.* Progress is being made on integrating the Vector Architecture model on the official gem5 repository [34]. It is possible to get an early access by cloning [35], a fork of the official gem5 repository that includes the Vector Architecture model. Note that this is a temporary repository only for early access, which will be removed as soon as the full code is hosted in the official gem5 repository.

## 4 RISC-V VECTORIZED BENCHMARK SUITE

The RISC-V Vectorized Benchmark Suite is a collection composed of seven data-parallel applications from different domains. The suite focuses on benchmarking vector micro-architectures;

Table 1. Vectorized Benchmark Suite Applications

| Application | Application Domain | Algoritdmical Model | DLP Pattern | Benchmark Suite |
|---|---|---|---|---|
| Blackscholes | Financial Analysis | Dense Linear Algebra | Regular | PARSEC/PARVEC |
| Canneal | Engineering | Unstructured Grids | Irregular | PARSEC/PARVEC |
| Jacobi-2D | Engineering | Dense Linear Algebra | Regular | PolyBench |
| Particle Filter | Medical Imaging | Structured Grids | Mix | Rodinia |
| Pathfinder | Grid Traversal | Dynamic Programming | Regular | Rodinia |
| Streamcluster | Data Mining | Dense Linear Algebra | Mix | PARSEC/PARVEC |
| Swaptions | Financial Analysis | MapReduce | Regular | PARSEC/PARVEC |

nevertheless, it can be used as well for Multimedia microarchitectures. Applications are Vector Length Agnostic; therefore, applications can be tested using short, medium, and large VLs. The current implementation is targeting RISC-V Architectures. It can be easily ported to any SIMD ISA, thanks to a wrapper library, which is developed to map vector intrinsics and math functions to the target architecture. A study was performed of different benchmarks to select the final applications for our suite, taking into account the following criteria:

*Applications from different domains.* Although the vector architectures have been traditionally applied to the supercomputing domain, this suite does not try to explore a single application domain, as was done by several benchmark suites.

*Applications with different data-level parallelism patterns.* Having different data-level parallelism patterns helps to test different real-world scenarios. While some vector architectures can take advantage of regular data-level parallelism patterns found in the application, these architectures could poorly execute another application that presents irregular data-level parallelism pattern. This property is interesting, enabling us to expose the weaknesses of some proposals/designs.

*Applications that cover most of the Vector ISA.* Finding an application, which uses almost all the vector ISA operations, is difficult. This is covered by selecting applications with different instruction uses. For example, applications were considered in which most of the instructions are memory operations or applications that are compute-bound, where arithmetic operations consume most of the execution time. Furthermore, vector ISAs typically feature a certain set of unique instructions such as slide or mask operations that are not found in scalar ISAs.

The RISC-V Vectorized Benchmark Suite is available to the computer architecture community to evaluate vector architecture designs. It is openly and temporarily available at *GitHub* [36]. Like the gem5 model, the suite will be hosted in the gem5 resources repository [37].

### 4.1 Vectorized Applications

Table 1 presents general information about the seven applications selected for the Vectorized Benchmark Suite. These applications can be categorized as having regular DLP, irregular DLP, or a mix of both. In the case of a mix there is, on the one hand, well-structured data accesses with regular and well-known address streams, including well-structured control flow corresponding to a regular DLP; and, on the other hand, less-structured data accesses with dynamic and difficult-to-predict address streams, and less-structured control flow representing irregular DLP [38].

Table 2 presents more detailed information for every application, such as the supported VL and the memory access pattern. It also indicates which applications stress the different vector microarchitecture modules such as the lane functional units or lane interconnection network. The final row shows if the application has intense communication with the scalar core, featuring a tight mixture of scalar and vector operations and accesses.

Table 2. Vectorized Benchmark Suite Applications

| Application | | Blackscholes | Canneal | Jacobi-2D | Particle Filter | Pathfinder | Stream cluster | Swaptions |
|---|---|---|---|---|---|---|---|---|
| Vector Length | Short | ✓ | ✓ | ✓ | ✓ | ✓ | ✓ | ✓ |
| | Medium | ✓ | ✓ | ✓ | ✓ | ✓ | | ✓ |
| | Large | ✓ | | ✓ | ✓ | ✓ | | ✓ |
| Memory Unit | Unit-stride | ✓ | | ✓ | ✓ | ✓ | | ✓ |
| | Indexed | | ✓ | | | | ✓ | |
| Vector Lane | Arithmetic | ✓ | ✓ | ✓ | ✓ | ✓ | ✓ | ✓ |
| | Mask | ✓ | | | ✓ | | ✓ | ✓ |
| Interconnection Network | Slides | | | ✓ | | ✓ | | |
| | Reductions | | ✓ | | | | ✓ | ✓ |
| Intensive Comm. with the Scalar core | | | ✓ | | ✓ | | ✓ | |

It is important to mention that several efforts are being made from the community to include the support for the new vector standard to the compiler. However, the current support is at an initial stage, limiting use to assembly code or intrinsics at best. In that sense, at this point, it is not possible to use auto-vectorization to obtain a different set of instructions. Having said that, all the applications (C and C++ programs) were extended by adding the RISC-V vectorized version. The implementations make use of intrinsics, meaning that the compiler will substitute the intrinsic by a sequence of predefined vector instructions. In that sense, the vector compiler only takes the decision to insert spill code (vector load/store) when the number of vector registers is not sufficient, and vector move instructions when a vector register is used as an argument in a function. Also, the code was written in a Vector Length Agnostic fashion, meaning that the same binary can be executed in different Vector Engine configurations with any modification. It is important to point out that a detailed description of the original applications is omitted; however, the suite in which the applications were taken from is specified. For all the applications, four input sets are available: small, medium, large and native, which are specified in the README file of the repository [36]. All results discussed in this document were performed using the large input set. The following subsections describe how the vectorized versions were implemented. Furthermore, the degree of the vectorization achieved and how it could lead to obtain some initial insights about expected performance is discussed.

*4.1.1 Blackscholes.* This application represents the wide field of analytic PDE solvers in general and their application in computational finance in particular. This application was taken from PARSEC, and a detailed description can be found in Reference [20].

Table 3 presents some statistics of the Blackscholes application for both the scalar and the vectorized implementations. The analysis for the Vectorized code takes into account three different MVL configurations: short-vectors (MVL=8 elements each 64-bit wide), medium-vectors (MVL=64 elements each 64-bit wide), and large-vectors (MVL=256 elements each 64-bit wide). *Total Instructions* represents the number of executed instructions (*Scalar Instructions + Total Vector Instructions*). *Scalar Instructions* represents only the instruction executed by the scalar core. *Total Vector* Instructions represents the instructions executed by the vector engine (*Vector Memory Instructions + Vector Arithmetic Instructions*). *Vector Operations* represents the number of operations performed by the vector instructions, while scalar instructions perform only one operation per instruction, vector instructions perform VL operations per vector instruction. *% of vectorization* is defined as the ratio of *Vector Operations* over the total number of operations (*Scalar Instructions + Vector Operations*). Thus, all previous data plus the *Average VL* give us an idea if most of the hardware resources will be used through the program execution. The same table structure is used for all applications in this study.

Table 3. Instruction-level Characterization of Blackscholes Application

|  | Scalar Code | Vectorized Code | | |
|---|---|---|---|---|
|  |  | MVL=8 elements each 64-bit | MVL=64 elements each 64-bit | MVL=256 elements each 64-bit |
| Total Instructions | 4,316,765,131 | 727,119,128 | 342,504,727 | 298,856,749 |
| Scalar Instructions | 4,316,765,131 | 484,635,928 | 312,194,327 | 291,279,149 |
| Vector Memory Instructions |  | 22,118,400 | 2,764,800 | 691,200 |
| Vector Arithmetic Instructions |  | 220,364,800 | 27,545,600 | 6,886,400 |
| Total Vector Instructions |  | 242,483,200 | 30,310,400 | 7,577,600 |
| Vector Operations |  | 1,939,865,600 | 1,939,865,600 | 1,939,865,600 |
| % of Vectorization |  | 80% | 86% | 87% |
| Average VL |  | 8 | 64 | 256 |

Blackscholes is a regular DLP application where there are no dependencies between each price computation. The runtime profile of the scalar code shows that around 12% of the total runtime is spent in the initialization phase, which is not vectorizable. This task executes 573,256,509 scalar instructions, which are not taken into account in the results presented in Table 3 to present the *% Vectorization* only of the region of interest (ROI). Around 85% of the runtime is used to compute the *BlkSchlsEqEuroNoDiv* and *CNDF functions,* which corresponds to the ROI. In this code section basic floating-point operations are computed, including *fadd, fsub, fmul, fdiv, fsqrt*, as well as logarithmic and exponential functions. The vectorization of these functions was straightforward, since it presents a very regular DLP. Furthermore, logarithmic and exponential functions where also vectorized, since most of the time is spent computing these functions.

The total number of instructions drops considerably for the vectorized versions, not only because one vector instruction represents many scalar operations of the same type. For example, a *vadd.v* instruction for a configuration of VL = 256 elements represents 256 scalar add instructions, but also it removes many control instructions needed to execute the desired number of operations. All the scalar instructions needed to write the "*for"* loop, or the data movement from/to memory produced by the limited number of physical scalar registers are removed. As the VL increases, the percentage of vectorization increases because of the ratio resulting from the number of vector operations, which remains equal, over the number of the total operations (scalar operations + vector operations), which decreases as a result of the reduction of the number scalar instructions, as shown in Table 3.

From previous data, it is possible to get an initial insight about the expected performance when executing the program in the vector architecture model. For example, obtaining the ratio between the number of the *Total instructions* of the serial version divided by the total number of operations (*Vector Operations + Scalar Instructions)* obtains a Vector-Accelerator-Only (VAO) speedup of 1.78x. Several factors can influence reducing or increasing this speedup. The execution of the remaining scalar instructions can be amortized underneath vector execution or the use of parallel lanes. Increasing the number of lanes would be expected to obtain a linear speedup increase, since it is a high DLP application. However, based on the fact that the number of *Vector Operations* are the same for all versions, and omitting the use of parallel lanes, is the VAO speedup of 1.78x going to be the same no matter the MVL configuration? The answer is no! One important difference between the scalar and the vector execution with one lane is that the execution of the vector operations can be pipelined, because all vector operations of one vector instruction are independent. So benefits from vectorization are two-fold: reduction in the total number of operations and faster execution of vector operations, thanks to pipelining. Then, as the MVL is increased, it is possible to hide the latency of individual operations. Then, as an initial conclusion, it can be said that for larger MVLs, it is possible to achieve higher speedups than 1.78x. As the number of lanes is increased, obtaining a linear speedup increase would be expected especially for larger MVL configurations where the *% of Vectorization* is higher.

*4.1.2 Canneal.* The Canneal application is focused on minimizing the routing cost of a chip design using cache-aware simulated annealing (SA). SA is a metaheuristic to approximate global optimization in a large search space for an optimization problem. This application is representative of engineering workloads and features fine-grained parallelism with lock-free synchronization and pseudo-random worst-case memory access pattern. This application was taken from PARSEC, and a detailed description can be found in Reference [20].

The candidate function to vectorize is *swap_cost,* which consumes most of the execution time. This is potentially a very vectorizable function, because it is composed of a couple of "*for*" loops in which three basic operations are performed: subtraction, absolute value, and addition. However, the data needed to perform these operations are the locations (coordinate x and y) of each input and output of the picked nodes; but these locations are not contiguous data in memory. The consequence is the need to use vector indexed load instructions, which are very costly operations. Furthermore, to use the vector indexed load operations, it is necessary to create the vector of pointers to each input/output element to have access to the pointer of the current element location. Creating it in every iteration is not good for the performance. Therefore, the class *netlist_elem* was expanded with a new array of pointers called *fan_locs* that stores the pointers to all inputs and outputs of each element; it is created in the initialization phase. Once the new array of pointers is added to the original code, the function *swap_cost* can be vectorized easily by first loading the *fan_locs* arrays of the picked nodes and then taking those loaded addresses to perform a couple of load indexed operations. Computing the routing cost is vectorized by using vector arithmetic and reduction instructions, sending the final result to the core to compute the final routing cost, and deciding if it should be swapped or not.

In this particular case, six configurations are presented in Table 4, since this application features some special case that will be discussed. Note that the initialization phase is not considered in the results presented in Table 4, to know the % Vectorization only of the region of interest. In Section 5.2, the impact in execution time caused by the addition of the *fan_locs* array in the initialization phase is shown.

There are three relevant observations about this application. First, there is no considerable reduction in the number of Scalar Instructions and Total Vector Instructions. This happens because the inputs and outputs for each node vary from 0 up to 22 connections for the large input set. Then, the largest requested VL in this application is 22 elements each, representing a short-vector application. From MVL=8 to MVL=32, there are slight variations in the number of Scalar Instructions and Total Vector Instructions, while from MVL=32 up to MVL=256, it remains constant. From MVL=8 to MVL=32 the variation in Total Vector Instructions is visible, since the "for" loops in the function *swap_cost* iterates 1, 2, 3, or up to 4 times, depending on the number of inputs and outputs for the picked node. This makes it possible to execute more Vector Instructions and, therefore, more Scalar Instructions, since to iterate several times causes the execution of more scalar instructions such as control instructions. For MVLs equal or larger than 32 elements, the number of Total Vector Instructions is the same. Since the larger requested VL is 22, hardware implementations with MVL=32 or bigger can operate on all the input and outputs for the current element in only one iteration.

Table 4. Instruction-level Characterization of Canneal Application

|  | Scalar Code | Vectorized Code | | |
| --- | --- | --- | --- | --- |
|  |  | MVL=8 elements each 64-bit | MVL=16 elements each 64-bit | MVL=32 elements each 64-bit |
| Total Instructions | 5,239,983,271 | 3,722,402,159 | 3,490,359,558 | 3,488,680,211 |
| Scalar Instructions | 5,239,983,271 | 3,368,424,160 | 3,218,719,265 | 3,217,635,854 |
| Vector Memory Instructions |  | 59,887,894 | 37,432,156 | 37,269,628 |
| Vector Arithmetic Instructions |  | 294,090,105 | 234,208,137 | 233,774,729 |
| Total Vector Instructions |  | 353,977,999 | 271,640,293 | 271,044,357 |
| Vector Operations |  | 2,450,191,462 | 3,102,641,472 | 4,078,370,559 |
| % of Vectorization |  | 42% | 49% | 56% |
| Average VL |  | 6.92 | 11.42 | 15.05 |
|  |  | MVL=64 elements each 64-bit | MVL=128 elements each 64-bit | MVL=256 elements each 64-bit |
| Total Instructions |  | 3,488,680,211 | 3,488,680,211 | 3,488,680,211 |
| Scalar Instructions |  | 3,217,635,854 | 3,217,635,854 | 3,217,635,854 |
| Vector Memory Instructions |  | 37,269,628 | 37,269,628 | 37,269,628 |
| Vector Arithmetic Instructions |  | 233,774,729 | 233,774,729 | 233,774,729 |
| Total Vector Instructions |  | 271,044,357 | 271,044,357 | 271,044,357 |
| Vector Operations |  | 6,030,736,943 | 9,926,999,575 | 17,727,994,975 |
| % of Vectorization |  | 65% | 76% | 85% |
| Average VL |  | 22.25 | 36.63 | 65.41 |

Second, the increase in the number of *Vector Operations* as the MVL is increased is substantial. Why does this occur? When analyzing the assembly code generated by the compiler, it is noted that there are two main factors that cause such an increase in the number of *Vector Operations*. First, there is code generated by the compiler when the number of vector registers is not sufficient. Then, additional vector load/store statements must be inserted to storing some registers in memory and restoring from memory to the original vector register when it finalizes. These extra statements are referred as *spill code*. The compiler knows nothing about the available MVL. In that sense, the compiler makes use of vector load/store instructions with the VL size equal to zero, which means to use the MVL that is supported by the micro-architecture. Then, the *Vector Operations* count increases as the MVL is increased. Second, every time that the function *swap_cost* is called, as part of the input arguments, there are three vector registers, which are a mask register and two vector registers holding the coordinates (x and y) for each node. The compiler creates vector move instructions to move those input arguments to specific registers, which can be used inside the vectorized function. This vector move instructions uses the MVL allowed by the hardware, meaning that the full vector is copied into a new one, no matter that only a small part of the vector register will be used, causing the increase in the number of *Vector Operations*.

Third, the *% of Vectorization* increases notably for large MVL configuration. This increase does not reflect what happens. It would be expected that the *% of Vectorization* increases as the MVL is increased because of the reduction of *Scalar Instructions*, while the number of vector operations remains constant. However, the number of *Vector Operations* also increases because of the different reasons explained above. Such an increase directly impacts the value of the *% of Vectorization*, since it is calculated as the ratio between the *Vector Operations over the Scalar Instructions + Vector Operations*. For that reason, this increase in the *% of Vectorization* does not provide a real improvement by vectorizing the application.

From previous data, it is possible to get an initial insight about the expected performance. In this case, it obtained a VAO speedup of 0.90x for the MVL=8. Unlike the Blackscholes application, Canneal presents intensive communication with the scalar core. When vector computation for every couple of picked nodes is finalized, the result is sent to the scalar core to evaluate if the nodes should be swapped or not. Then, it can be expected that many of the remaining scalar instructions cannot be amortized underneath vector execution. Contrary to Blackscholes, as the MVL is increased, a speedup decrease is expected, since the number of *Vector Operations* increases notably.

*4.1.3 Jacobi-2D.* Jacobi-2D is an iterative algorithm for determining the solutions of a diagonally dominant system of linear equations. This solver is often used in computational science as part of scientific and engineering applications. This application was taken from PolyBench, and a detailed description can be found in Reference [19].

The Jacobi solver is a very interesting application, because it can be vectorized by using vector arithmetic operations, vector memory instructions, and vector element manipulation instructions. In that sense, not only the Lanes and the Vector Memory Unit are evaluated, but also the interconnection between the lanes. These vector element manipulation instructions are *vslide1up.v* and *vslide1down.v*, which move elements one position up and down a vector register. It is possible to load a fraction of one row and operate on it by applying vslide1up.v to obtain the left neighboring nodes and *vslide1down.v* to obtain the right neighboring nodes. Once the left and right neighboring nodes are aligned and top and bottom neighboring nodes are loaded, it is possible to operate on them in parallel.

One interesting observation is the slight variation in the number of *Vector Operations*. This happens because a nested "for" loop is implemented to go through the matrix to compute one complete iteration. Outside the "for" loop, a vector instruction is defined, which sets a constant. This instruction is executed only once per iteration, and for the large input set, 4,000 iterations are defined, meaning that this instruction is executed 4,000 times. For MVL=8 elements, means that in one iteration, this instruction is going to execute 8 *Vector Operations* (32,000 *Vector Operations* after finalizing all the iterations). On the contrary, for MVL=256 elements, means that in each iteration, this instruction is going to execute 256 *Vector Operations* (1,024,000 *Vector Operations* after finalizing all the iterations). Then, if those values are subtracted from the total *Vector Operations* in the corresponding configuration, this will result in the same number of *Vector Operations* for all the cases.

As an initial insight based on the data presented in Table 5, it is possible to obtain a VAO speedup of 1.09x. Additionally, for larger MVLs it is possible to achieve higher speedups. As increasing the number of lanes obtains a linear speedup, an increase would be expected, especially for larger MVL configurations where the *% of Vectorization* is higher.

*4.1.4 Particle Filter.* The Particle Filter (PF) application is a statistical estimator of the location of a target object given noisy measurements of the state. In image analysis, the PF is utilized widely in applications such as facial recognition. This particular implementation is optimized for tracking cells, particularly leukocytes (white blood cells). This application was taken from Rodinia, and a detailed description can be found in Reference [22]. This application uses special operations with masks, which send resultant data to the scalar core and were not used in previous applications. These vector instructions are *vfirst.m*, which finds the lowest-numbered active element of the source mask vector that has its least-significant bit set and writes that element's index to a scalar register; and *vpopc.m*, which counts the number of mask elements of the active elements of the vector source mask register that have their least-significant bit set and writes the result to a scalar register. Also, this application uses complex operations such as logarithm, cosine, and square root.

Table 5. Instruction-level Characterization of Jacobi-2D Application

|  | Scalar Code | Vectorized Code | | |
| --- | --- | --- | --- | --- |
|  |  | MVL=8 elements each 64-bit | MVL=64 elements each 64-bit | MVL=256 elements each 64-bit |
| Total Instructions | 4,797,698,032 | 1,665,765,868 | 328,373,875 | 185,081,872 |
| Scalar Instructions | 4,797,698,032 | 1,275,617,868 | 279,601,875 | 172,885,872 |
| Vector Memory Instructions |  | 65,280,000 | 8,160,000 | 2,040,000 |
| Vector Arithmetic Instructions |  | 259,894,400 | 32,489,600 | 8,124,800 |
| Vector Elem Manipulation Inst |  | 64,973,600 | 8,122,400 | 2,031,200 |
| Total Vector Instructions |  | 390,148,000 | 48,772,000 | 12,196,000 |
| Vector Operations |  | 3,121,184,000 | 3,121,408,000 | 3,122,176,000 |
| % of Vectorization |  | 71% | 92% | 95% |
| Average VL |  | 8 | 64 | 256 |

The task in charge of applying a predefined motion model that represents the estimated path that the object will follow is a good candidate to be vectorized, because the same operations are applied to all objects in the frame. Furthermore, to apply the motion model, it is necessary to generate a sequence of random numbers using the Box-Muller transformation, which makes use of expensive operations such as logarithm, cosine, and square root. The task that consumes most execution time is the guess updates based on the current location of the object. These new guesses are used by the following frame in the video to iterate again. This task is implemented in a nested for loop, which performs a sequential search returning an index value to update the arrays *arrayX* and *arrayY*. This task can be implemented by first using a vector comparison instruction to obtain a mask representing the active elements for that iteration. Later, the *vfirst.m* instruction is used to know if there is at least one active element in the generated mask and its corresponding position. When the criteria are met, the position of each active element is obtained to finally use the *vpopc.m* instruction to check if all elements in the vector have been set, breaking the inner loop. Otherwise, the program continues with a new iteration until all elements get the updated position.

Table 6 presents some statistics for the PF application. This application presents a very high percentage of vectorization increasing as the MVL is increased, achieving up to 91% of vectorization. However, the number of instructions for an MVL=8 is reduced by 4x, and as the MVL is increased, the number of instructions decreases up to 16x. The slight variation in the number of *Vector Operations* is similar to the case presented in Jacobi-2D. Instructions outside of the nested loops are responsible for this slight variation.

As was the case for previous applications, it is possible to get an initial insight about the expected performance. By analyzing PF, it is possible to get VAO speedup of 1.27x for the MVL=8 and up to 10.16x for eight-lane configuration. Unlike Blackscholes and Jacobi-2D applications, which are regular and high DLP applications, Particle Filter presents a mix of DLP. Nevertheless, by using *vfirst.m* and *vpopc.m* vector instructions, it is possible to handle the less-structured control flow in a relatively easy way. For sure, this will cause some implications in terms of performance, since both instructions cause intensive communication with the scalar core. It is expected that many of the remaining scalar instructions cannot be amortized underneath vector execution.

Table 6. Instruction-level Characterization of Particle Filter Application

|  | Scalar Code | Vectorized Code | | |
| --- | --- | --- | --- | --- |
|  |  | MVL=8 elements each 64-bit | MVL=64 elements each 64-bit | MVL=256 elements each 64-bit |
| Total Instructions | 20,232,505,095 | 4,993,215,636 | 1,617,632,096 | 1,260,531,622 |
| Scalar Instructions | 20,232,505,095 | 3,446,128,079 | 1,423,641,027 | 1,211,546,181 |
| Vector Memory Instructions |  | 1,607,712 | 200,992 | 50,272 |
| Vector Arithmetic Instructions |  | 1,545,479,845 | 193,790,077 | 48,935,169 |
| Total Vector Instructions |  | 1,547,087,557 | 193,991,069 | 48,985,441 |
| Vector Operations |  | 12,376,700,456 | 12,415,428,416 | 12,540,272,896 |
| % of Vectorization |  | 78% | 90% | 91% |
| Average VL |  | 8 | 64 | 256 |

Table 7. Instruction-level Characterization of Pathfinder Application

|  | Scalar Code | Vectorized Code | | |
| --- | --- | --- | --- | --- |
|  |  | MVL=8 elements each 64-bit | MVL=64 elements each 64-bit | MVL=256 elements each 64-bit |
| Total Instructions | 6,213,455,512 | 1,337,948,580 | 402,094,500 | 301,824,392 |
| Scalar Instructions | 6,213,455,512 | 1,037,138,340 | 364,493,220 | 292,424,072 |
| Vector Memory Instructions |  | 100,270,080 | 12,533,760 | 3,133,440 |
| Vector Arithmetic Instructions |  | 120,324,096 | 15,040,512 | 3,760,128 |
| Vector Ele Manipulation Instr |  | 80,216,064 | 10027008 | 2,506,752 |
| Total Vector Instructions |  | 300,810,240 | 37,601,280 | 9,400,320 |
| Vector Operations |  | 2,406,481,920 | 2,406,481,920 | 2,406,481,920 |
| % of Vectorization |  | 70% | 87% | 89% |
| Average VL |  | 8 | 64 | 256 |

*4.1.5 Pathfinder.* The Pathfinder application uses the ghost zone optimization technique to find the shortest paths on a 2-D grid. This application was taken from Rodinia, and a detailed description can be found in Reference [22]. The reason to select this application is due to the high percentage of vector element manipulation instructions.

One interesting aspect of this application is that the algorithm implemented to find the shortest paths inside *run* function is composed of a nested loop. For each node, comparisons with its corresponding neighboring nodes are performed to find the smallest weight and adding it to the current node weight. This task is easily implemented using the vector *slide1up* and *slide1down* operations to accommodate the neighboring nodes in the same position and operate on it to finally store the resultant data. Vector element manipulation instructions reported in Table 7 consume 26% of the total vector instructions, meaning that the lane interconnection can be stressed by this application. The number of instructions for MVL=8 is reduced by 4.6x, and as the MVL is increased, the number of instructions decreases up to 20.5x. However, the percentage of vectorization increases as the MVL is increased, achieving up to 89% of vectorization.

Table 8. Instruction-level Characterization of Streamcluster Application

| | Scalar Code | Vectorized Code | | |
| --- | --- | --- | --- | --- |
| | | MVL=8 elements each 64-bit | MVL=64 elements each 64-bit | MVL=128 elements each 64-bit |
| Total Instructions | 36,068,326,139 | 6,349,730,434 | 2,599,142,070 | 2,331,242,835 |
| Scalar Instructions | 36,068,326,139 | 4,325,602,994 | 2,241,943,122 | 2,093,110,203 |
| Vector Memory Instructions | | 952,530,560 | 119,066,316 | 59,533,158 |
| Vector Arithmetic Instructions | | 1,071,596,880 | 238,132,632 | 178,599,474 |
| Total Vector Instructions | | 2,024,127,440 | 357,198,948 | 238,132,632 |
| Vector Operations | | 16,193,019,520 | 22,860,732,672 | 30,480,976,896 |
| % of Vectorization | | 79% | 91% | 94% |
| Average VL | | 8 | 64 | 128 |

As initial insight for Pathfinder application based on the data presented in Table 7, it is possible to obtain a VAO speedup of 1.8x for an MVL=8 and one-lane configuration. Also, it would be expected to obtain a linear increase as the number of lanes increase, achieving up to 14.4x for eight-lane configuration. Finally, for larger MVLs it is possible to achieve higher speedups than 1.8x.

*4.1.6 Streamcluster.* The Streamcluster application solves an online clustering problem. For a stream of input points, it finds a predetermined number of medians so each point is assigned to its nearest center. The program is memory bound for low-dimensional data and becomes increasingly computationally intensive as the dimensionality increases. This application was taken from PARSEC, and a detailed description can be found in Reference [20].

The candidate function to vectorize is "*dist,*" which consumes most of the execution time (95%). This function computes the squared Euclidian distance between two points, which is calculated as the cumulative addition of the distances between each of the point's dimensions. This function is highly vectorizable, but the number of vector arithmetic operations is almost the same as the memory operations needed in each iteration, which means that a vector implementation could be limited by the memory subsystem. Then, it is implemented by using two vector loads and two vector arithmetic operations. Outside the inner "for" loop, a vector reduction is needed to get the cumulative addition. The resultant scalar value of the reduction is sent immediately to the scalar core to compute the cost of opening a new center. This is done by using the *vfirst.m* instruction. Note that this last step could cause a huge impact on performance, since before starting a new iteration, it is necessary to evaluate if the cost of opening a new center would be advantageous. This computation is made by the scalar core, meaning that for every iteration, the vector engine receives a block of instructions, computes them, and returns a scalar value. Later the vector engine will be idle while the scalar core evaluates the results.

Table 8 presents the instruction-level characterization for Streamcluster. The number of instructions for an MVL=8 is reduced by 5.6x, and as the VL is increased, the number of instructions decreases by up to 15.4x. However, the percentage of vectorization increases correspondingly with the MVL, achieving up to 94% of vectorization. One interesting aspect of this application is that the variation in the number of vector operations for every MVL configuration is considerable. This variation in the number of *Vector Operations* is similar to the case presented in Jacobi-2D and Particle Filter; instructions outside the nested loops cause this variation. However, in this case the variation is huge, because the number of instructions outside the loop are half of the total instructions in one call to "dist" function. Then, the number of *Vector Operations* is equal to the number of

Table 9. Instruction-level Characterization of Swaptions Application

|  | Scalar Code | Vectorized Code | | |
| --- | --- | --- | --- | --- |
|  |  | MVL=8 elements each 64-bit | MVL=64 elements each 64-bit | MVL=256 elements each 64-bit |
| Total Instructions | 26,846,776,223 | 6,337,441,159 | 1,022,467,455 | 456,078,412 |
| Scalar Instructions | 26,846,776,223 | 4,173,151,623 | 751,931,263 | 388,444,364 |
| Vector Memory Instructions |  | 370,323,456 | 46,290,432 | 11,572,608 |
| Vector Arithmetic Instructions |  | 1,793,966,080 | 224,245,760 | 56,061,440 |
| Total Vector Instructions |  | 2,164,289,536 | 270,536,192 | 67,634,048 |
| Vector Operations |  | 17,314,316,288 | 17,314,316,288 | 17,314,316,288 |
| % of Vectorization |  | 81% | 96% | 98% |
| Average VL |  | 8 | 64 | 256 |

*Vector Instructions* multiplied by the MVL parameter, meaning that the larger the MVL is, the more vector operations are executed. As a preliminary observation, it can be said that it is an application that does not benefit much from larger MVLs, because more vector operations are executed for this case compared with shorter MVLs.

Based on the data presented in Table 8, it is possible to get an initial insight about the expected performance. A VAO speedup of 1.75x for an MVL=8 and one lane configuration could be achieved. As the MVL increases, it is not clear if speedup improvements could be expected, since the number of *Vector Operations* also increases. The increase in the number of parallel lanes could give a slight speedup increase. As mentioned before, this application is memory bound for the large input set, then, the speedup could be limited by the memory subsystem. This discussion continues in Section 5.7, showing the results of the application executed on different Vector engine configurations.

*4.1.7 Swaptions.* The Swaptions application uses the Heath-Jarrow-Morton (HJM) framework to price a portfolio of swaptions based on Monte Carlo simulation to compute the prices. The HJM framework describes how interest rates evolve for risk management and asset-liability management for a class of models. This application was taken from PARSEC, and a detailed description can be found in Reference [20].

The application has three most time-consuming functions, which are *RanUnif, serialB,* and *CumNormalInv. RanUnif* function is in charge of performing a random initialization. This initialization consumes nearly 10% of the total execution time. This function can be vectorized by defining MVL number of seeds instead of only one as in the scalar version. By vectorizing this function, the output differs by a very small difference. It is because the following generated random numbers are calculated based on the new vector of seeds instead of only one. The final standard error is sometimes slightly smaller or slightly bigger. The next function is *serialB*, which generates the cumulative normal distribution matrix necessary to calculate the HJM paths. It consumes 26% of the total execution time. This function is implemented using three nested "*for*" loops. Inside this nested "*for*" loop, the function *CumNormalInv* is called to compute the inverse of the cumulative normal distribution function. This function consumes 22.98% of the total execution time. This function has a regular DLP pattern and computes basic floating-point operations, including *fadd, fsub, fmul*, and *fdiv*, and also logarithmic functions.

Table 9 presents the statistics of the Swaptions application. The number of instructions for MVL=8 is reduced by 4.2x, and as the MVL is increased, the number of instructions decreases up to 74x. The percentage of vectorization increases as the MVL is increased, achieving up to 98% of vectorization. Considering previous data, it is possible to achieve a VAO speedup of 1.24x. This application presents a regular and high DLP pattern. For this reason, as the number of lanes increase a linear speedup increase would be expected especially for larger MVL configurations where the *% of Vectorization* is higher.

Table 10. Gem5 Evaluation Environment

| Config. 1 to 4 | Config. 5 to 8 | Config. 9 to 12 | Config. 13 to 16 | Config. 17 to 20 | Config. 21 to 24 |
|---|---|---|---|---|---|
| **Scalar Core** - Clock Frequency - 2 GHz | | | | | |
| Dual-Issue 64-bit RISC-V superscalar in-order pipeline, | | | | | |
| **Vector Engine** - Clock Frequency - 1 GHz | | | | | |
| **# Lanes** | # Lanes | # Lanes | # Lanes | # Lanes | # Lanes |
| 1 \| 2 \| 4 \| 8 | 1 \| 2 \| 4 \| 8 | 1 \| 2 \| 4 \| 8 | 1 \| 2 \| 4 \| 8 | 1 \| 2 \| 4 \| 8 | 1 \| 2 \| 4 \| 8 |
| MVL 512-bit ($8 \times 64$-bit) | MVL 1024-bit ($16 \times 64$-bit) | MVL 2048-bit ($32 \times 64$-bit) | MVL 4096-bit ($64 \times 64$-bit) | MVL 8192-bit ($128 \times 64$-bit) | MVL 16384-bit ($256 \times 64$-bit) |
| Renaming with 40 Physical Registers | | | | | |
| VRF **2.5 KB** | VRF 5 KB | VRF 10 KB | VRF 20 KB | VRF 40 KB | VRF 80 KB |
| 1 pipelined arithmetic unit / Lane | | | | | |
| VMU with 1 Memory Port connected to L2, 512-bit memory interface | | | | | |
| Ring topology for Lane Interconnection | | | | | |
| **Memory System** | | | | | |
| 32 KB L1I – latency 4 cycles – cache line 512-bit | | | | | |
| 32 KB L1D – latency 4 cycles – cache line 512-bit | | | | | |
| 256 KB L2 – latency 12 cycles – cache line 512-bit | | | | | |
| 2 GB DDR3 Memory | | | | | |

## 5 EVALUATION

This work is intended to be used as a base model for research on vector architectures. Rather than presenting approaches for the best performance, a discussion is presented that evaluates the results obtained in the analysis presented in Section 4, and the results are obtained when the vectorized RISC-V benchmarks are executed in the previously presented gem5 simulator with several vector engine configurations. The vector engine is attached to a superscalar in-order processor, with the system configurations shown in Table 10. Twenty-four configurations are evaluated for the vector engine. First, from one up to eight lanes was configured. By doing this, and setting only one memory port, it could be enough to feed up to eight lanes, taking into account that the cache line size is set to 512-bit (8 elements each 64-bit), and with every cache line request it is possible to send one element to each lane in an interleaved fashion. The MVL allowed varies from 512-bits up to 16,384-bits. All the configurations implement renaming with 40 physical registers, leading to VRF sizes from 2.5 KB to 80 KB for the different configurations. Issue queues with in-order issue logic are set. Each lane features only one pipelined arithmetic unit. Also, a ring topology for lane interconnection is chosen. The designer is able to choose simpler or more aggressive configurations according to the research requirements.

### 5.1 Blackscholes

Figure 4 shows the execution time (left axis) and speedup (right axis) obtained for the different configurations. Also, the VAO speedup (right axis) is shown to discuss these results. The obtained speedup for MVL=8 and one lane configuration is 2.22x, which is higher than the VAO speedup (1.78x). This happens because many of the remaining scalar instructions can be amortized underneath vector execution. As the MVL is increased, speedup improvements are seen as discussed in Section 4.1.1. Regarding the expected linear increase as the number of lanes is increased, it is not completely accurate. Configurations with small- and medium-size MVL do not benefit considerably from adding more lanes, unlike configurations that use large vectors. This is mainly because, in all configurations, the *start-up time* is incurred. As mentioned before, the *start-up time* is principally determined by the pipeline latency of the vector functional unit. However, the number of read ports in the VRF can also influence the *start-up time*. For all the configurations, a VRF with only one read/write port was set, meaning that to feed the source buffers with the corresponding source operands, one, two, or three cycles are needed, depending on the number of sources used by the instruction. Then, those extra cycles are added to the *start-up time*. For low MVL configurations, the *start-up time* is high compared to the total execution time of the instruction. In this case, the advantage for regular and high DLP applications for large MVL is visible since the *start-up time* becomes minimal compared with the total execution time of the instruction.

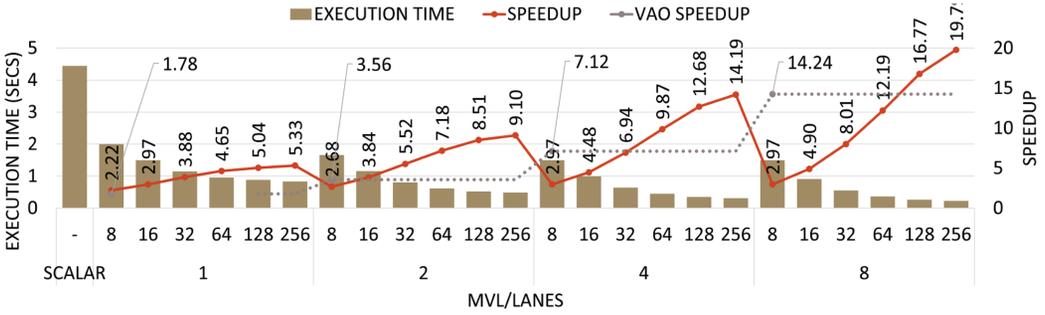

Fig. 4. Blackscholes runtime/speedup.

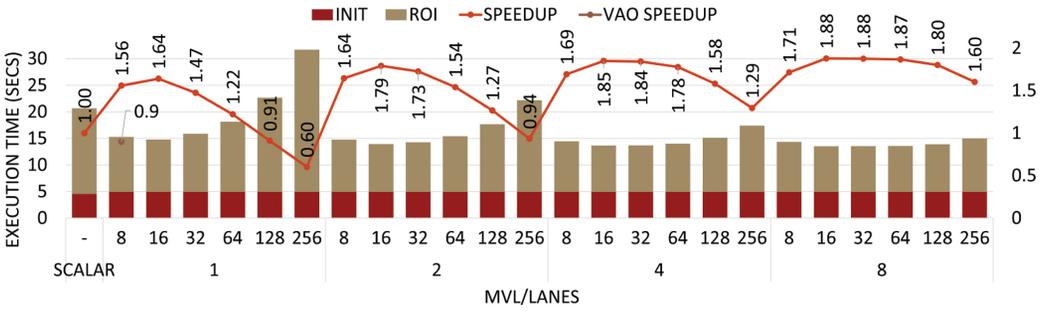

Fig. 5. Canneal runtime/speedup.

## 5.2 Canneal

Contrary to the previous application (Blackscholes), which benefited from any MVL because of its high DLP, Canneal represents shorter vectors. As presented in Table 1, this is an irregular DLP application, which increases the complexity to improve the performance even for the vectorized version. Although the analysis presented in Section 4.1.2, mentions that the largest VL for this application is 22 elements, the application was executed with all the configurations. This is done to show the behavior when applications with short vectors are executed in hardware for large vectors. In addition, Figure 5 shows the initialization phase (INIT), which increases from 4.56 seconds in the scalar version to 4.95 seconds in the vectorized version by adding the array *fan_locs*. ROI represents the region of interest to be evaluated.

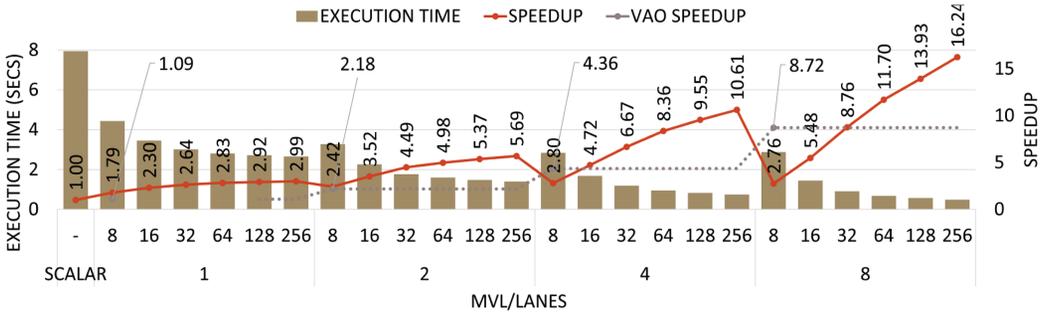

Fig. 6. Jacobi-2D runtime/speedup.

As was pointed out in Section 4.1.2, the configuration with MVL=8 elements would obtain a VAO speedup of 0.9x, and this speedup would decrease as the MVL parameter is increased. Results presented in Figure 5 exhibit a behavior close to that expected. In this case, the configuration with an MVL=16 obtained the best performance, achieving 1.64x of speedup over the scalar version (ROI Section) for the single lane configuration and 1.88x of speedup for the eight-lane configuration, and as the MVL parameter was increased the speedup starts to decrease, as is expected. This difference with the VAO speedup suggests that about half of the remaining scalar instructions can be amortized beneath vector execution. Note that the VAO speedup for the different lane configurations is not shown, since, for this application, a linear increase is not expected. In general, the low speedup is mainly because this application has an irregular DLP pattern, including intensive indexed memory accesses, which are very expensive in terms of latency. Furthermore, performance is limited by the large number of scalar instructions executed, and as mentioned before, about half of scalar instructions cannot be amortized underneath vector execution. This is because the scalar core waits for the result from the vector engine to compute the final routing cost and decides if the current elements should be swapped or not. For larger MVL configurations, the speedup starts to decrease, as was expected. In fact, for MVL=128 and 256, the scalar version is faster than the vectorized version, since the number of Vector operations increase notably because of the complementary instructions added by the compiler, and which uses the MVL allowed by the hardware. In Section 4.1.2 two factors were identified causing the possible speedup degradation, the context save and context restore task, and the generated vector move instructions necessary to use the vector registers as arguments in the function *swap_cost*. One more culprit that cannot be inferred from analyzing the assembly code. According to the specifications [12], when the application VL is smaller than the MVL, the remaining elements (tail elements) must be set as "0." This means that for larger MVL, more tail elements must be set. On applications where the MVL is fully used, this case is not a problem.

### 5.3 Jacobi-2D

Results presented in Figure 6 show a speedup of 1.79x for one-lane configuration, which is higher than the VAO speedup (1.09x). This happens because most of the remaining scalar instructions can be amortized beneath vector execution. As the MVL increases, the speedup increases up to 2.99x, as was expected. Regarding the increase in the number of lanes and the expected linear increase, it is not completely accurate. Configurations with small- and medium-size MVL do not benefit considerably from adding more lanes, unlike configurations that use large vectors. Looking at the configurations with an MVL=256, almost a linear speedup increase can be seen. As mentioned before, the speedup increase is strongly related to the *start-up time*. It is incurred in all configurations, but for larger MVL configurations, the *start-up time* becomes negligible.

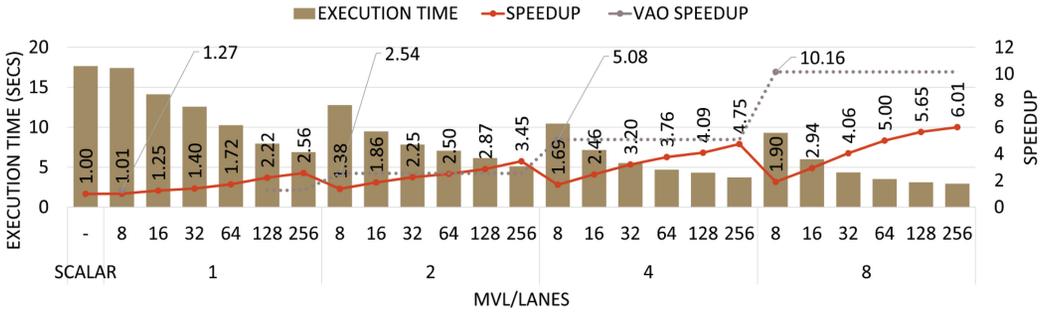

Fig. 7. Particle filter runtime/speedup.

## 5.4 Particle Filter

Particle Filter (PF) is an interesting application to analyze, because it combines the use of expensive operations like logarithm, cosine, and square root with special operations with masks. *vfirst.m* and *vpopc.m* vector mask instructions write the final results to a scalar register. In that sense, these operations cause the scalar core to stall because of the higher number of scalar dependencies.

According to the static code analysis presented in Section 4.1.4, the VAO speedup is 1.27x for MVL=8 and one-lane configuration. There is no speedup over the scalar version, as can be seen in Figure 7. In general, all the VAO speedups are higher than those already obtained for the different configurations. As previously suggested, the final speedup would be affected by a considerable number of stalls in the scalar core, which would not be removed until the vector engine finishes the computation of the current iteration. Based on the results, it can be concluded that for applications such as Particle Filter in which many of the remaining scalar instructions cannot be amortized beneath vector execution because of the generated scalar dependencies, there could be significant improvements by using an out-of-order superscalar core instead of a superscalar in-order core. For the out-of-order case, it would be possible to advance independent scalar instructions and also continue feeding the vector engine.

## 5.5 Pathfinder

As mentioned in Section 4.1.5, Pathfinder application is interesting, since it presents a high percentage of vector element manipulation instructions. Thus, it is possible to evaluate the implemented interconnection topology between lanes. In this case, the base model is using a ring interconnection, where to move one element to another lane, the cost in latency is the distance between the origin and the destination lanes. Several elements can be computed in parallel in multiple lanes. In this particular case, the algorithm makes use of slide1up and slide1down vector instructions, where the elements are displaced by only one position. In that sense, the ring interconnection is enough to get a good speedup for this application, since it will require only one cycle to move one element from the current lane to the destination lane. Also, each lane can send one element to the ring interconnection in the same cycle, and one cycle later all the lanes will receive their corresponding data. It is clear that these operations can take advantage of the parallel lanes.

The results presented in Figure 8 exhibit a behavior very close to the VAO speedup (1.8x) for MVL=8 and one lane configuration, and as the MVL parameter increases, higher speedups can be achieved. For multi-lane configurations, the expected linear speedup increase is not achieved. It is clear that configurations with small- and medium-size MVL do not benefit considerably from adding more lanes, unlike configurations that use large vectors. As mentioned several times before, this is because, in all configurations, the start-up time has to be paid, but for larger MVLs, the start-up time becomes negligible.

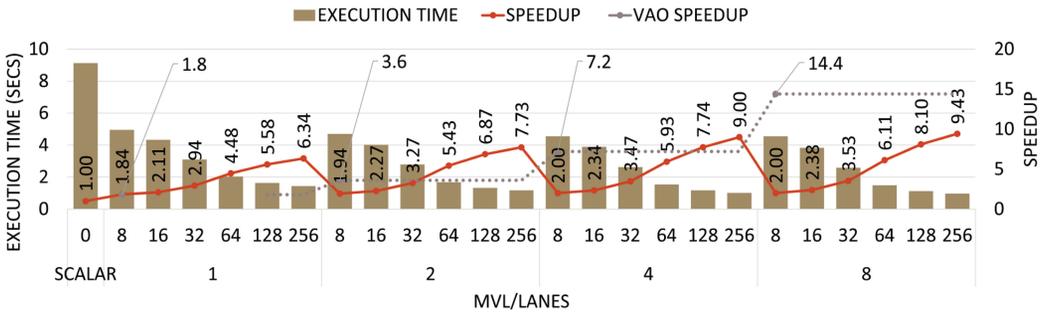

Fig. 8. Pathfinder runtime/speedup.

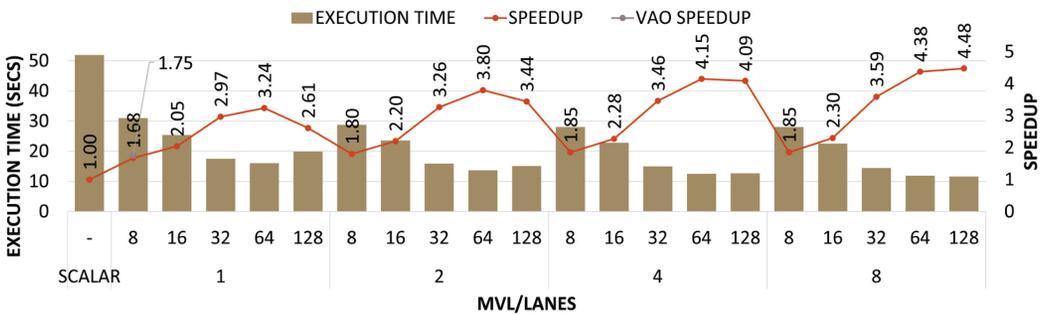

Fig. 9. Streamcluster runtime/speedup.

### 5.6 Streamcluster

Results are presented in Figure 9. For MVL=8 and one lane configuration, a speedup of 1.68x is obtained, which is very close to the VAO speedup (1.75). As the MVL slightly increases, improvements can be seen. However, from MVL=64 to MVL=128, the speedup decreases notably. There are several factors causing this low speedup increase and the sudden decrease when MVL=128. First, for larger vector lengths, the number of vector operations increases notably, then, more time is needed to execute those vector operations. Second, there is a reduction operation after the "*for*" loop. This operation is executed only once regardless of the VL size. Then, for short vectors, this operation has relatively less overhead; but for long vectors, this operation consumes more time. Finally, the resultant scalar value of the reduction is sent immediately to the scalar core to compute the cost of opening a new center. Thus, in this step, the core stalls until it receives these data, then computes the cost and finally iterates again.

The addition of parallel lanes helps to achieve better speedups. The improvements are not good enough to justify the use of parallel lanes for this application. As mentioned before, this application is memory bound, and then the speedup is mainly limited by the memory subsystem.

### 5.7 Swaptions

For the scalar implementation, a block size of 32 elements presents the best performance. For a larger block size, the L1 cache miss rate increases. For the vectorized version, it is a little bit different, as a block size of 128 elements gives better performance, achieving 6.8x speedup over the scalar version regardless of the cache miss increase. The largest VL allowed by the application is related to the block size parameter (*BLOCK_SIZE*); when increasing the *BLOCK_SIZE* parameter beyond 128, the speedup decreases, as shown in Figure 10 (SPEEDUP *L2-256KB*). It is because the miss rate in the last level cache (LLC) starts to increase. For those vector architecture designs targeting large vectors, the size of the LLC is crucial, because a miss in LLC will require access to the main memory. In Reference [20] a miss rate study is presented for all programs in the PARSEC benchmark suite, where for Swaptions, it is shown that for a cache size of 1 MB, the miss rate is reduced notably.

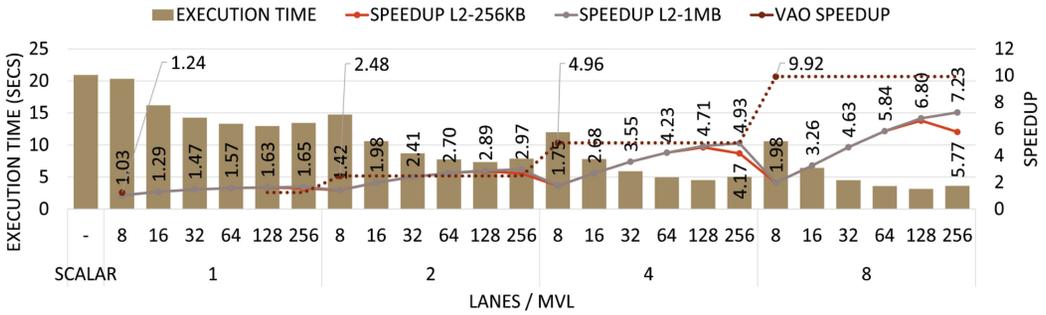

Fig. 10. Swaptions runtime/speedup.

Figure 10 also shows a new evaluation for Swaptions (*SPEEDUP L2-1MB*), changing the L2 Cache size in the system configuration presented in Table 10 from 256 KB to 1 MB. The results support the observations presented in Reference [20]. It is possible to observe that for an MVL configuration smaller than 128, the improvements are not visible (but there are some improvements). For MVL configuration larger or equal to 128, the improvements can be seen, achieving a speedup of 7.23x for an MVL=256. In contrast, in previous results, for MVL equal to 256, the performance is decreased. This is an example where it is important to design a vector architecture with an optimal MVL based on the predefined cache configuration.

It is interesting to compare the obtained versus the expected results. The obtained speedup for MVL=8 and one lane configuration is 1.03x, which is lower than the VAO speedup (1.24x). Configurations with small- and medium-size MVL do not benefit considerably from adding more lanes, unlike configurations that use large vectors. This is mainly because, in all configurations, the *start-up time* is incurred. Contrarily, for MVL=256 and the different number of lanes, almost a linear speedup increase can be seen.

## 6 RELATED WORK

Stanic [39] presents a set of tools for rapid initial research on vector architectures. The first tool is called VALib, a library that enables hand-crafted vectorization of applications by adding calls, which is similar to programming using intrinsics. VALib is not bound to any specific vector ISA. By using this tool, it is possible to collect data for detailed instruction-level characterization and to generate input traces for a second tool called SimpleVector. This second tool is a fast trace-driven simulator used to estimate the execution time of a vectorized application on a candidate vector micro-architecture. This simulator can be used for preliminary evaluation and early parameter exploration but does not provide the accuracy given by execution-driven simulators.

Cebrian [23, 40] presents PARVEC, a vectorized version of the PARSEC benchmark suite. ParVec vectorized 8 of the 13 applications of the PARSEC suite for SSE, AVX, and NEON ISA's, Some of them obtaining high speedup over the scalar implementation (Blackscholes, Swaptions) and others (Fluidanimate, Vips) did not get any speedup improvement mainly because of the nature of the application (organization of the input, not related to size, etc.). The ParVec suite is available for the computer architecture community. The lack of a micro-architectural simulator for those ISA's does not allow the computer architecture community to test new ideas at the micro-architectural level. Although the ISA's supported by PARVEC can be classified as Multimedia instruction set extensions, the similarities with the code for RISC-V Vector Extension ISA is extensive, mainly for arithmetic and covert operations. However, there are others like slide operations that usually are not presented in short vector ISA's. In this sense, PARVEC was a great tool for understanding how to vectorize some applications from the PARSEC Benchmark Suite.

The ARM Architecture research team has been working on tools for the community to boost the use of ARM infrastructure in academia and they have presented these tools in several talks about Vector Architecture for HPC based on Arm SVE [41]. The SVE tool-suite includes the Arm Compiler, the Arm Instruction Emulator, and the Research Enablement kit, which allows system modeling using gem5. The implemented gem5 models correspond to the Armv8-A-based CPU timing model (HPI) with support for SVE. The toolkit also includes documentation about how to run the benchmarks, specifically the PARSEC benchmark suite.

## 7 CONCLUSIONS

This work has presented two tools. First, an extended version of the gem5 simulator that includes a RISC-V Vector Architecture model. This model can be configured with different parameters (MVL, number of physical registers, number of lanes, etc.), having a flexible and customizable model that fits with the research requirements. Second, a RISC-V Vectorized Benchmark Suite. In addition, a study of every vectorized application and its corresponding execution in the vector engine model is given, highlighting the degree of vectorization achieved with the applications and the close relationship with the expected and obtained performance.

Some future directions our team is going to follow are given as the next steps in this research, taking as a base all the previously described work. On the one hand, it is well-known that applications such as Blackscholes, Jacobi-2D, and Swaptions, which present regular and high DLP patterns, will exploit in an efficient way designs for short, medium, and large MVL. On the other hand, applications that present low DLP can only exploit in an efficient way hardware designed for short vectors. However, executing those applications in hardware for large MVL brings several disadvantages, as was shown in the study. In that sense, finding a way to obtain a more general vector architecture able to handle different DLP patterns in an efficient way is a challenge in examination. Reconfigurable vector architectures with the ability to modify their own structures depending on the application's needs could improve the performance for those applications where most of the hardware remains unused. For example, it is possible to make independent use of different lanes by advancing the execution of instructions in each one, however, what would happen if the data required by one lane were allocated in a different lane? Perhaps it is necessary to devise a new control structure to keep track of the current location of every physical register, to issue the instructions in a more intelligent way and avoid data movement as much as possible. Another option could be to use all unused space in the VRF as a stack, and then the vector engine could perform a context save in a faster way. All these ideas and many more could be studied using this new platform, providing a useful and functional tool for the computer architecture community.

A RISC-V Simulator and Benchmark Suite 38:29